\documentstyle[12pt]{article}

\newcommand{\postscript}[2]{\vspace 5cm}

\renewcommand{\Large}{\large}

\begin{document}

\def\eB{ $\!^8B$ }
\def\sBe{ $\!^7Be$ }

\newcommand{\beq}{\begin{equation}}
\newcommand{\eeq}{\end{equation}}
\newcommand{\beqa}{\begin{eqnarray}}
\newcommand{\eeqa}{\end{eqnarray}}
\def\lag{Lagrangian}
\def\ra{\rightarrow}
\def\x{\times}
\def\etal{{\it et al., }}
\newcommand{\ie}{{\em i.e., }}
\def\sinn{\sin^2 \theta_W}
\newcommand{\conn}[4]{ {\bf #1}, #2, #3 (19#4)}
\newcommand{\RMP}[3]{{\em Rev. Mod. Phys.} {\bf #1}, #2 (19#3)}
\newcommand{\PR}[3]{{\em Phys. Rev.} {\bf #1}, #2 (19#3)}
\newcommand{\PL}[3]{{\em Phys. Lett.} {\bf #1}, #2 (19#3)}
\newcommand{\Rep}[3]{{\em Phys. Rep.} {\bf #1}, #2 (19#3)}
\newcommand{\Ann}[3]{{\em Ann. Rev. Nucl. Sci.} {\bf #1}, #2
(19#3)}
\newcommand{\NS}[3]{{\em Nucl. Sci.} {\bf #1}, #2 (19#3)}
\newcommand{\PRL}[3]{{\em Phys. Rev. Lett.} {\bf #1}, #2 (19#3)}
\newcommand{\NP}[3]{{\em Nucl. Phys.} {\bf #1}, #2 (19#3)}
\newcommand{\con}[3]{{\bf #1}, #2 (19#3)}
\def\kev{\; {\rm keV} }
\def\mev{\; {\rm MeV} }
\def\ev{\; {\rm eV} }
\def\tev{\; {\rm TeV} }
\def\eV{\; {\rm eV} }
\def\gev{\; {\rm GeV} }
\def\etc{ {\it etc}}

\begin{center}

{\Large \bf SOLAR NEUTRINOS}\footnote{Invited talk presented
at {\it 32nd International School of Subnuclear Physics}, Erice,
July 1994.}

\vspace{3ex}

PAUL LANGACKER \\ {\it University of Pennsylvania \\ Department of Physics
\\ Philadelphia, Pennsylvania, USA 19104-6396}
\\ UPR-0640T, \ \ \ \today

\vspace{4ex}

Abstract

\end{center}

The status of the solar neutrino problem is reviewed.  Attempts to explain
the observed deficit and spectral distortion, both by astrophysical and
particle physics methods, are described.  It is argued that the comparison
of all experiments strongly prefers the particle physics solutions.


\section{Introduction}

There is no direct evidence for neutrino mass from any laboratory
experiment.  Nevertheless, there are strong hints from theory,
astrophysics, and cosmology that the neutrinos may have small masses.  It
has long been known that the observed high energy solar neutrinos produced
in \eB\ decays are suppressed by a factor of two to three compared to
expectations of the standard solar model.  This by itself could be
accounted for by astrophysical or nuclear physics effects.  However, there
are now four experiments of three types, which are sensitive to different
parts of the spectrum.  By comparing the experiments it is possible to
infer that most of the suppression is in the middle of the spectrum,
associated with the \sBe\ line and the lower energy part of the \eB\
$\beta$ spectrum.  This is inconsistent with any known astrophysical or
nuclear physics explanation, and strongly suggests new neutrino properties,
such as MSW matter-enhanced neutrino oscillations.  This conclusion can be
reached by using any two of the three types of existing experiments.
Future experiments will yield considerable new
information.  By searching for spectral distortions,
day/night effects, and anomalous ratios of neutral current to charged
current events it should be possible to either confirm or falsify the need
for neutrino oscillations, independent of astrophysical uncertainties.

\section{Neutrino Mass}

There is no compelling laboratory evidence for non-zero neutrino mass.  The
direct limits from kinematic searches for the masses yield the upper limits
\cite{pdg}
\beqa m_{\nu_e} &<& 5.1 \ev, \;{\rm tritium} \; \beta \; {\rm
decay}\protect\footnotemark \nonumber \\
m_{\nu_\mu} &<& 270 \kev, \pi \ra \mu\nu_\mu \label{eq1} \\
m_{\nu_\tau} &<& 31 \mev, \tau \ra \nu_\tau + n\pi.\nonumber
\eeqa
\protect\footnotetext{The tritium $\beta$ decay experiments all yield
negative $m^2$ values, with a weighted average
$m_{\nu_e}^2 = (-96 \pm 21)\,\ {\rm eV}^2$,
suggesting a common systematic or theoretical uncertainty in the
experiments.  Until this is understood the precise upper limit must be
considered somewhat questionable.}

On the other hand, most extensions of the standard predict non-zero masses
at some level \cite{nurev}.  Unified theories and other extended gauge
groups with a large mass scale $M_{\rm new}$ often predict a seesaw-type mass
$m_\nu \sim v^2/M_{new}$, where $v = 246$~GeV is the weak scale.  Many
other models with Higgs triplets or loops involving new Higgs particles
also generate neutrino mass at some level.  There are also several hints
for non-zero mass from astrophysics and cosmology.  All four solar neutrino
experiments, Homestake \cite{Homestake}, Kamiokande \cite{Kam}, SAGE
\cite{sage}, and GALLEX \cite{gallex}, observe a deficit of neutrinos
compared to the standard solar model expectation \cite{Bahcall}, suggesting
that there is either nonstandard astrophysics or new neutrino properties.
However, the relative rates observed by Homestake and Kamiokande yield a
simple measure of the distortion of the spectral shape, suggesting that
most of the suppression is in the middle of the spectrum, {\it i.e.}, of
the
\sBe\ line and probably the low energy part of the \eB\ spectrum, with less
suppression at lower and higher energies \cite{suppression}--\cite{ssd}.
This cannot
be accounted for by any known astrophysical or nuclear physics explanation
and strongly suggests new neutrino properties.  More recently, the two
gallium experiments, SAGE and GALLEX, have accumulated reasonably good
statistics.  Their rates are low compared to any reasonable solar model,
standard or nonstandard, again suggesting the need for new neutrino
properties.  One can infer the suppression of the \sBe\ line from any two
types of experiment, gallium/Kamiokande, gallium/Homestake, or
Homestake/Kamiokande.

Amongst the new neutrino properties the Mikheyev-Smirnov-Wolfenstein (MSW)
\cite{msw} mechanism of matter-enhanced neutrino oscillations explains the
data very well, and does suppress the middle of the spectrum for some
regions of neutrino mass and mixings \cite{hl2} -- \cite{ber}.  Although
it remains to be verified by future experiments, the MSW mechanism
is quite promising
and would be extremely exciting for particle physics.  Solar neutrinos are
also useful for
astronomy.  Either with or without the MSW effect one will be able to use
them to probe the different components of the neutrino
spectrum \cite{modind} and therefore to do astronomy on the solar core.
Even with present data, if one accepts the 2-flavor MSW interpretation
the neutrino parameters and the temperature
$T_C$ of the core of the sun
can be simultaneously determined
\cite{modind}, yielding $T_C/T_{SSM} = 1.00
\pm 0.03$, where $T_{SSM} = 1.6 \x 10^7$ K is the prediction of the standard
solar model \cite{BPSSM,TCSSM}.  Similarly, one can use the
present data, again assuming MSW, to simultaneously determine the flux of
\eB\ neutrinos.  One finds \cite{modind} $\phi (\,^8B)/\phi_{SSM} = 1.15
\pm 0.53$ for the flux relative to the standard solar model prediction.

Another hint of neutrino mass is the anomalous ratio
$\nu_\mu/\nu_e$ suggested by underground searches for neutrinos produced  by
interactions of cosmic rays in the atmosphere \cite{atmun}.

Finally, the combination of COBE data \cite{COBE} and the distribution of
galaxies on large and small scales is hard to understand on the basis of
simple cold dark matter.  One possibility is that in addition to cold dark
matter there is a small admixture \cite{mixed} of hot dark matter,
presumably due to a massive $\tau$ neutrino with a mass in the range
$m_{\nu_\tau} \sim (1 - 35) \ev$ \cite{blud}.  There are, however,
alternative explanations, such as a cosmological constant, topological
defects, or a tilted initial spectrum.

If one accepts some or all of these hints for neutrino mass, the typical
range $(10^{-3} -$~few~$)\ev$ is consistent with the general seesaw-type
scale \cite{seesaw,BKL} with new physics around $M_{new} \sim
10^{12}\, \gev$.  This is a typical value that might be expected for the
heavy neutrino mass scale in supersymmetric grand unification \cite{cl}.

\section{Solar Neutrinos}

The basic energy source in our sun is believed to be the $pp$ cycle, in
which four protons fuse to form $\!^4He$, {\it i.e.}, $4p \ra \,^4He + 2e^+
+ 2 \nu_e$ \cite{Bahcall}.  The dominant initial reactions are
\beqa pp &\ra& \,^2H + e^+ + \nu_e \nonumber \\
\,^2H+p &\ra & \,^3 He + \gamma.\label{eq2} \eeqa
The first of these results in the low energy $pp$~neutrinos.  Their number
is the firmest prediction of the solar model because it is closely tied to
the overall luminosity.  However, they are very hard to detect because of
their low energy.  Most of the $\!^4He$ is from
\beq \,^3He + \,^3He \ra \,^4He + 2p.\label{eq3} \eeq
However, approximately 15\% is believed to be produced from the sequence
\beqa \,^3 He + \,^4He &\ra & \,^7Be + \gamma \nonumber \\
e^- + \,^7Be &\ra& \,^7Li + \nu_e \label{eq4} \\
\,^7 Li + p &\ra& 2 \,^4He,\nonumber \eeqa
which yields $\,^7Be$ neutrinos at two discrete energies, one of which is
somewhat above the $pp$ spectrum.  Finally, a rare side reaction,
\beqa p + \,^7Be & \ra & \,^8B + \gamma \nonumber \\
\,^8B &\ra& \,^8Be^* + e^+ + \nu_e \label{eq5} \\
\,^8Be^* &\ra & 2 \,^4He \nonumber \eeqa
is associated with about 0.02\% of the produced $\!^4He$.  This is
insignificant energetically, but the resulting \eB\ neutrino spectrum
extends to much higher energy than the others, so they are easier to
detect.  The predicted spectrum \cite{Bahcall} from the $pp$ cycle and the
rarer CNO cycle neutrinos are shown in Figure~\ref{fig1}.

\begin{figure}
\postscript{/home/pgl/fort/nc/graph/solar/ssm_spectrum.ps}{0.85} 
\caption[]{Predicted spectrum of solar neutrinos.}
\label{fig1}
\end{figure}

\begin{table} \centering
\begin{tabular}{|lcrc|} \hline \hline
Experiment & reaction & threshold & location \\ \hline
Kamiokande & $\nu_ee \ra \nu_ee$ & 7.3 MeV & Japan \\
Homestake  & $\nu_e + \,^{37}Cl \ra e^- + \,^{37}Ar $ & 0.8 MeV & United
States \\
SAGE & $\nu_e \,^{71}Ga \ra e^+ + \,^{71}Ge $ & 0.233 MeV & Russia \\
GALLEX & $\nu_e \,^{71}Ga \ra e^+ + \,^{71}Ge $ & 0.233 MeV & Italy \\
\hline \hline
\end{tabular}
\caption[]{Presently operating solar neutrino experiments.}
\label{tab1}
\end{table}

There are currently four solar neutrino experiments, as shown in
Table~\ref{tab1}.  The Kamiokande experiment \cite{Kam} is a 1 KT water
Cerenkov detector which measures the energy of the produced electrons.  It
is only sensitive to the highest energy \eB\ neutrinos, but it is a real
time experiment.  It also yields some information on the direction of the
incident neutrinos, which allowed Kamiokande to show that the neutrinos are
really coming from the sun.  Homestake \cite{Homestake} was the first solar
neutrino experiment, and it has been running for 25 years.  It
consists of $10^5$ gallons of $C_2 Cl_4$, and detects neutrinos via capture
on the chlorine.  It has a much lower energy threshold than Kamiokande, and
is therefore sensitive to the higher \sBe\ line as well as the lower energy
parts of the \eB\ spectrum.  However, its largest sensitivity is still to
higher energies.  In the last few years two gallium experiments, the SAGE
experiment  in the Baksan Neutrino observatory in
the Caucasus mountains,
and the GALLEX experiment in the Gran Sasso tunnel in
Italy, have been running.  They are sensitive to the low energy $pp$
neutrinos, as well as to the higher energy neutrinos.  The predicted
contributions to the gallium and chlorine experiments in the standard solar
model are shown in Table~\ref{tab2}.

\begin{table} \centering
\begin{tabular}{|lcr|} \hline \hline
Neutrino & \ & \ \\
Source   & Homestake & Gallium \\
$pp$  & 0.0 & 70.8 \\
$pep$ & 0.2 & 3.1 \\
\sBe\ & 1.2 & 35.8 \\
\eB\ & 6.2 & 13.8 \\
$\,^{13}N$ & 0.1 & 3.0 \\
$\,^{15}O$ & 0.3 & 4.9 \\
Total & 8.0 $\pm$ 3.0 & $131.5^{+21}_{-17}$ \\ \hline \hline
\end{tabular}
\caption[]{Predicted rates in SNU
($10^{-36}$ atom$^{-1} \;s^{-1}$)
from the various flux components for the
chlorine and gallium experiments, from \protect\cite{BPSSM}.  The
uncertainties are the total theoretical range, $\sim 3\sigma$.}
\label{tab2} \end{table}

The results of the experiments are compared with the predictions of two
standard solar models, that of Bahcall and Pinsonneault (BP) \cite{BPSSM}
and that of Turck-Chieze and Lopes (TCL) \cite{TCSSM}, in Table~\ref{tab3}.
\begin{table} \centering  \footnotesize 
\begin{tabular}{|lccccc|} \hline \hline
Exp & BP SSM & TCL SSM & Exp & Exp/BP & Exp/TCL \\ \hline
Kamiokande & $5.69 \pm 0.82$ & $ 4.4 \pm 1.1$ & $2.89^{+0.22}_{-0.21} \pm
0.35$ & $0.50 \pm 0.07 [0.07]$ & $0.65 \pm 0.09 [0.16] $ \\
Homestake & $8 \pm 1$ & $6.4 \pm 1.4$ & $2.55 \pm 0.17 \pm 0.18$ & $0.32
\pm 0.03 [0.04] $ & $0.40 \pm 0.04 [ 0.09]$ \\
Gallium & $131.5^{+7}_{-6}$ & $122.5 \pm 7$ & $77\pm 9$ & $0.59 \pm 0.07
[0.03]$ & $0.63 \pm 0.07 [0.04]$ \\
(combined) & \            & \ & \ & \ & \ \\
SAGE & \ & \ & $74^{+13\,+5}_{-12\,-7}$ & \ & \\
GALLEX & \ & \ & $79 \pm 10 \pm 6       $ & \ & \\ \hline \hline
\end{tabular}
\caption[]{Predictions of the BP and TCL standard solar models for the
Kamiokande, Homestake, and Gallium experiments compared with the
experimental rates.  The Kamiokande flux is in units of $10^6/cm^2 \ s$, while
the Homestake and gallium rates are in SNU.  The experimental rates
relative to the theoretical predictions are shown in the last two
columns, where the first uncertainty is experimental and the second is
theoretical.  After 1986 the Homestake rate was slightly higher $2.76 \pm
0.31$~SNU, which corresponds to $0.35 \pm 0.04[0.04]$ compared to BP and
$0.43 \pm 0.05 [0.09]$ compared to TCL. All uncertainties are 1 $\sigma$. }
\label{tab3}
\end{table}

It is seen that the predictions for the Kamiokande and Homestake
experiments are between $1/3$ and $1/2$ of the BP expectations; the
Kamiokande rate is still low compared to TCL, although somewhat closer; the
Gallium rates are about 60\% of the predictions.  This deficit of neutrinos
is shown in Figure~\ref{fig2}, which also displays the typical neutrino
energy to which each class of experiment is sensitive.
\begin{figure}
\postscript{/home/pgl/fort/nc/graph/solar/snp_bpssm.ps}{0.65}
\postscript{/home/pgl/fort/nc/graph/solar/snp_tcssm.ps}{0.65}
\caption[]{The experimental observations relative to the predictions of the
BP and TCL standard solar models.  The error bars on the points are
experimental; the (1 $\sigma$)
theoretical uncertainties are displayed separately.
Each experiment is sensitive to a range of neutrino energies. The values
shown represent typical energies for each experiment.}
\label{fig2}
\end{figure}

The solar neutrino problem has two aspects.  The older and less significant
is that all of the experiments are below the SSM predictions.  This was
never a serious concern for the Kamiokande and Homestake
experiments individually, which
are mainly sensitive to the high energy \eB\ neutrinos which are less
reliably predicted.  However, the predictions for the gallium experiments
are harder to modify due to the constraint of the overall solar luminosity,
and the statistics on the gallium experiments are starting to be good
enough that the deficit observed there is hard to account for.

A second and more serious problem is that the Kamiokande rate indicates
less suppression than the Homestake rate.  The Homestake experiment has a
lower energy threshold, and the lower observed rate suggests that there is
more suppression in the middle of the spectrum (the \sBe\ line and the
lower energy part of the \eB\ spectrum) than at higher energies.  This is
very hard to account for by astrophysical or nuclear physics mechanisms: the
\eB\ is made from \sBe\ (eqn (\ref{eq5})), so any suppression of \sBe\
should be accompanied by at least as much suppression of \eB. Furthermore,
all known
mechanisms for distorting the \eB\  $\beta$ decay spectrum are negligible
\cite{astdistort}.

There are several generic explanations of the solar neutrino problem.  In
discussing astrophysical/nuclear solutions, one must distinguish between
the uncertainties in the standard solar models, and nonstandard solar
models with new physics ingredients.  A second possibility is particle
physics solutions, which invoke nonstandard neutrino properties.  Of these I
will concentrate on what I consider the simplest and most favored
explanation, the Mikheyev-Smirnov-Wolfenstein (MSW) matter enhanced
conversion of one neutrino flavor into another \cite{msw}.  There are
other possible explanations, such as the more complicated 3-flavor MSW
\cite{msw3}, vacuum oscillations \cite{vac1}--\cite{h1},
neutrino decay
\cite{decay}, large magnetic moments \cite{mag}, or violation of the
equivalence principle \cite{equiv}.  Many of these are disfavored by the
data and are, to my mind, less natural.  The third possibility is that some
or all of the experiments are wrong.  However, this is becoming harder to
accept, because the same difficulties follow from any two of the
classes of experiments:  one no longer has to believe all of the results to
conclude that there is a problem.

\section{Astrophysical Solutions}

Attempts to account for the observations by astrophysical and nuclear
physics explanations have to be divided into two categories.  First let us
consider the standard solar models \cite{BPSSM,TCSSM}.  I will
use as an example the Bahcall-Pinnsonneault model \cite{BPSSM}, which
includes helium diffusion and an improved estimate of $S_{17}$, which is an
energy dependent factor$^{2,3}$\footnotetext[2]{Similarly $S_{11}$,
$S_{33}$, and $S_{34}$
are proportional to $\sigma(pp \ra \,^2He^+ \nu_e), \sigma
(\,^3He + \,^3He \ra \,^4He + 2p)$, and $\sigma (\,^3He + \,^4He \ra \,^7Be
+ \gamma)$, respectively.}
\footnotetext[3]{This cross section is especially uncertain because two
measurements are not in agreement \protect\cite{Bahcall}.}
proportional to $\sigma (p + \,^7Be \ra \gamma + \,^8B)$.  The BP model is
in agreement with most other calculations when the same inputs are used,
and is in agreement with helioseismology data and information about main
sequence stars.  Within the model there are uncertainties due to the input
parameters.  In particular, uncertainties in the metallicity, $Z$, and
other contributions to uncertainties in the opacities are important.  They
mainly manifest themselves for the neutrinos by modifying the predicted
core temperature, $T_C$, to which the predicted rates of higher energy
neutrinos are extremely sensitive.  There are also nuclear cross section
uncertainties, both for the production reactions within the sun and for the
detectors.  The production cross sections
are problematic because the energies involved are lower
than can easily be measured in the laboratory. The experimental cross
sections must therefore be extrapolated to low energy, and, since they involve
barrier penetration, there is considerable energy dependence.
Nevertheless, given the
canonical estimates of the uncertainties the standard solar model is certainly
excluded by the data.

What is still possible, however, are nonstandard solar models (NSSM).
These involve new ingredients compared to the SSM, {\it
e.g.,} there could be new physics inputs such as core rotation, magnetic
fields, WIMP's, or gravitational settlings.  Most of these
effects manifest themselves by leading to a cooler sun.  The high
energy neutrinos are very sensitive to this; one estimate
\cite{bu} is that the fluxes vary as $\phi (B) \sim T^{18}_C$ and $\phi
(Be) \sim T^8_C$, so that small reductions in the core temperature could
suppress the number of high energy neutrinos significantly.  It should be
cautioned that such nonstandard models may conflict with helioseismology
data, main sequence stairs, {\it etc}.  I will not worry about that, but
will simply concentrate on whether they can, in fact, describe the neutrino
data.

Another type of nonstandard model is one in which there are large
differences in the cross sections from what is usually assumed.  In
particular, it is possible that $S_{17}$ is lower than the usual estimates,
and one preliminary experiment suggests that that may be the case
\cite{gal}.  This would certainly lower the predicted flux of \eB\
neutrinos, and could easily account for the Kamiokande results.  However,
it does not explain the larger suppression of Homestake compared to
Kamiokande, and, in fact, it aggravates that difficulty since most of the
expected Homestake rate is also from \eB\ neutrinos.

\subsection{Cool Sun Models}

\addtocounter{footnote}{2}

There are a number of ways to analyze these possibilities.  Many of the
nonstandard models manifest themselves for neutrino production by leading
to a cooler solar core.\footnote{The relation of cool sun models to
specific nonstandard models is discussed in \protect\cite{hbl,castell}.}
This may ultimately  be due to changes in the metallicity, the opacities, or
the nuclear cross sections.  The fluxes of high energy neutrinos are
especially sensitive to the core temperature, mainly because of the energy
dependence of the cross sections.  We have carried out an analysis to see
whether the data can be described by a lower temperature \cite{bkl2} --
\cite{thesis}.  We start by assuming the Bahcall-Ulrich estimates \cite{bu}
of the temperature dependence of the \eB\ and \sBe\ fluxes, namely $\phi
(B) \sim T^{18}_C$ and $\phi(Be) \sim T^8_C$, where $T_C$ is the temperature
of the solar core in units of the SSM prediction, $ 1.6 \x 10^7$ K. For
small deviations from the standard model $\phi(pp)$ is predicted to vary as
$T_C^{-1.2}$, {\it i.e.}, the $pp$ rate increases.  This is
necessary to maintain ${\cal L} \sim$~constant, where ${\cal L}$
is the solar luminosity.  However, we will be using the power laws to
describe larger departures from the standard solar model than the usual
estimates, so instead we choose $\phi(pp)$ so that the
luminosity is constant. {\it I.e.}, one knows that the basic energy
mechanism is the conversion
\beq 4p \ra \,^4He + 2e^+ + 2 \nu_e + 26.7 \;\mev . \label{eq6} \eeq
Assuming that the sun is quasi-static, the overall luminosity tells us the
total energy generation rate, and this in turn puts a constraint on the
total neutrino fluxes, namely
\beq \phi (pp)+ \phi(pep) + 0.958 \phi (Be) + 0.955 \phi {\rm(CNO)} \; =
6.57 \x 10^{10} {\rm cm}^{-2}s^{-1}, \label{eq7} \eeq
where the coefficients correct for the neutrino energies.
We will use this to constrain\footnote{The $pep$ neutrinos are a minor
component from the reaction $p + e^- + p \ra  \,^2H + \nu_e$.  They are
strongly correlated with the $pp$ flux in most nonstandard models.}
     $\phi(pp)$. One then has
\beqa R_{Cl} &=& 0.32 \pm 0.03 = (1 \pm 0.033) \left[ 0.775 (1 \pm 0.089)
T_C^{18} \right. \nonumber \\
\, &\, & \;\;\;\;\;\;\;\;\;\;\; \left. +
0.150 (1 \pm 0.036) T^8_C + {\rm small} \right]    \label{eq8} \\
R_{Kam} &=& 0.50 \pm 0.07 = (1 \pm 0.089) T^{18}_C  \label{eq9} \\
R_{Ga} &=& 0.59 \pm 0.07 = (1 \pm 0.04) \left[ 0.538
\frac{\phi(pp)}{\phi_{\rm SSM}} \right. \nonumber \\
&& \;\;\;\;\; \left. 0.105 (1 \pm 0.089) T^{18}_C + 0.272 (1 \pm 0.036)
T^8_C + {\rm small} \right],\label{eq10} \eeqa
where $R$ is the rate and $T_C$ is the
temperature, both relative to the standard solar model prediction.  The
additional uncertainties in the formulas are from nuclear cross sections,
for the production rates within the sun and for the detector
cross sections, which have to be properly correlated between experiments.
``Small'' represents the contributions of minor flux components ($pep$ and
CNO).  The results are insensitive to how these are treated.
It is apparent that it is impossible to describe the data for any value of
$T_C$.  The best fit is nominally $T_C = 0.93 \pm 0.01$, but it has a
terrible $\chi^2$ of $ 15.7$ for 2~df, which is excluded at 99.96\%~cl.  The
problem is that the different experiments require different temperatures,
namely $0.96 \pm 0.01$, $0.92 \pm 0.01$ and $0.71 \pm 0.14$ for Kamiokande,
Homestake, and the combined gallium results, respectively.

The specific power laws, taken from Bahcall and Ulrich \cite{bu}, were based
on the standard solar model, while here we are extrapolating to nonstandard
models.  In fact one obtains equally strong conclusions for essentially any
temperature exponents, provided only that the \eB\ neutrinos are more
temperature sensitive than the \sBe\ neutrinos.  Since the \sBe\ neutrinos
are made first, this is certainly a reasonable assumption.  One finds
similar conclusions even if the $S_{17}$ error is increased significantly.
In fact, the conclusion is strengthened if there is a smaller $S_{17}$,
because the Kamiokande rate is then in better agreement with the
SSM, leaving less room to vary the temperature to account for the chlorine
experiment.  Thus, it seems that the cool sun models are not an explanation
of the data.

One can generalize  to more general fits, in which not only the
core temperature but the nuclear cross sections are allowed to vary.  For
example, Dearborn, Shi, and Schramm \cite{ssd}
have considered the case in which
$T_C$, $S_{17}$, and $S_{34}$ are all free parameters ($S_{34}$ is
proportional to $\sigma (\,^4He + \,^3He \ra \,^7Be + \gamma)$).  Again,
they come to the conclusion that one cannot account for the data within
reasonable ranges for these parameters.

\subsection{Model Independent Analysis}
\label{modindsec}

We decided some time ago to analyze the data in as model independent a
context as possible \cite{hbl,mod2}.  Though most explicitly-constructed
nonstandard models involve either the temperature or the cross sections
\cite{hbl,castell} there is always the possibility of very nonstandard
physical inputs which cannot be described in this way.  The idea in a model
independent analysis is that all that really matters for the neutrinos are
the magnitudes $\phi(pp)$, $\phi (Be)$, $\phi (B)$ and $\phi(CNO)$ of the
various flux components.  We can then analyze the data making only three
minimal assumptions.  One is that the solar luminosity is quasi-static and
generated by the normal nuclear fusion reactions.  This leads\footnote{The
luminosity observed now corresponds to the energy that was generated in the
core $10^4$ yr. ago.  The quasi-static assumption allows one to equate the
present luminosity with the present energy production rate.  One can
actually relax this assumption and come to essentially the same
conclusion.} to the constraints (\ref{eq6}) and (\ref{eq7}).  The second
assumption is that astrophysical mechanisms cannot distort the shape of the
\eB\ spectrum significantly from what is given by normal weak interactions.
Nobody has found any astrophysical mechanism that can significantly distort
the shape, and all explicitly studied mechanisms are negligibly small
\cite{astdistort}.  It is this assumption which differentiates
astrophysical mechanisms from MSW, which can distort the shape
significantly.  Our third assumption is that the experiments are correct, as
are the detector cross section calculations.

In this (almost) most general possible solar model all one has to play with
are the four neutrino flux components\footnote{The uncertainties associated
with $\phi (pep)$ are negligible.} subject to the luminosity
constraint.  The strategy is to fit the data to the \sBe\ and \eB\ fluxes.
For each set of fluxes, one varies $\phi (pp)$ and $\phi(CNO)$ so as to
get the best fit.  The CNO and other minor fluxes play little role because
they are bounded below by zero, and larger values make the fits worse.  The
constraints from the individual classes of experiments are shown in
Figure~\ref{fig3}.
\begin{figure}
\postscript{/home/pgl/fort/nc/graph/solar/fff_1.ps}{0.6}
\caption[]{\protect\sBe\ and \protect\eB\ fluxes relative to the standard
solar model prediction as constrained by different classes of experiments.
The SSM corresponds to the point ($1,1$) while the uncertainties in the SSM
are shown as an ellipse. From~\cite{thesis}.}
\label{fig3}
\end{figure}

Figure~\ref{fig4} displays the allowed region from all data.  The best fit
would occur in the unphysical region of negative \sBe\ fluxes.
Constraining the flux to be positive, the best fit requires $\,^7Be < 7\%$
and $\,^8B =41 \pm 4\%$ of the SSM \cite{hbl,modind}.  This, however, has a
poor $\chi^2$.  One finds $\chi^2_{\rm min} = 3.3$ for 1~d.f.,~which is
excluded at 93\% CL, \ie  it is only marginally
allowed statistically.

More important, the best fit it is in a region that is hard to account
for by astrophysical mechanisms.  Figure~\protect\ref{fig4} also
displays predictions of the BP and TCL standard solar models, the 1,000
Monte Carlos SSMs of Bahcall and Ulrich (dots) \cite{bu}, other explicitly
constructed nonstandard models \cite{nssm}, and the general predictions of
cool sun and low $S_{17}$ models.

Most of the nonstandard models are approximately parameterized by the cool
sun models \cite{hbl,castell}, but none come close to what is required by
the data.  As can be seen in the figure, the low $S_{17}$ models are
especially far from the observation.  The problem is that the data is
requiring an almost total suppression of \sBe\ neutrinos compared to the
\eB\ neutrinos \cite{suppression}--\cite{ssd}.
That is hard to understand astrophysically,
because boron is produced from the beryllium by proton capture.  If one
gets rid of all of the beryllium there is no plausible explanation of why
so {\em much} \eB\ is still produced.

People have occasionally questioned the validity of the Homestake results,
although there is no clear reason to doubt them.  In fact, the data is now
sufficiently good that one can draw the same conclusion about the complete
suppression of \sBe\ neutrinos from any two types of experiments, as can be
seen in Table~\ref{tab4}.  For example, Figure~\ref{fig5} shows the
constraint if the chlorine data is omitted.  In this case the overall
$\chi^2$ is acceptable, but the allowed region is still not consistent with
any explicit solar model.
One concludes that it is unlikely that any NSSM will explain the data
unless at least two of the experiments are wrong \cite{modind,jbhmn,dfl}.

One can reach much the same conclusion in another way.  In
Figure~\protect\ref{fig5b} the predictions for gallium are shown for various
explicitly constructed nonstandard models which agree with Kamiokande but
ignore the Homestake rate.  All of the explicit models predict rates in
excess of 100 SNU, well above the combined observations.

\begin{figure}
\postscript{/home/pgl/fort/nc/graph/solar/beb_curr_all.ps}{0.6}
\caption[]{90\% CL combined fit for the \sBe\ and \eB\ fluxes.  The best fit,
which occurs at $\phi(\,^7Be) < 7$\% and $\phi(\,^8B) = 0.41 \pm 0.04$
relative to the SSM, has a poor $\chi^2$ of 3.3 for 1 d.f.  Also shown are
the predictions of the BP and TCL SSM's, 1000 Monte Carlo SSM's
\protect\cite{bu}, various nonstandard solar models, and the models
characterized by a low $T_C$ or low $S_{17}$. From
\protect\cite{hbl,modind,thesis}.}
\label{fig4}
\end{figure}

\begin{figure}
\postscript{/home/pgl/fort/nc/graph/solar/beb_curr_no-cl.ps}{0.6}
\caption[]{Constraints on the \eB\  and \sBe\ fluxes without the Homestake
data.}
\label{fig5}
\end{figure}

\begin{table} \centering
\begin{tabular}{|lcccc|} \hline \hline
       & $pp$  & \protect\sBe\ & \protect\eB\ & CNO \\ \hline
Without MSW: & \ &     \         &   \         &    \ \\
Kam $+Cl + Ga$ & $1.089 - 1.095 $ & $<0.07$ & $0.41 \pm 0.04$ & $< 0.26$ \\
Kam $+Cl     $ & $1.084 - 1.095 $ & $<0.13$ & $0.42 \pm 0.04$ & $< 0.38$ \\
Kam $+Ga     $ & $1.085 - 1.095 $ & $<0.13$ & $0.50 \pm 0.07$ & $< 0.56$ \\
    $Cl + Ga $ & $1.082 - 1.095 $ & $<0.16$ & $0.38 \pm 0.05$ & $< 0.72$ \\
\hline
With    MSW: & \ &     \         &   \         &    \ \\
Kam $+Cl + Ga$ & $<1.095$ & -- & $1.15 \pm 0.53$ & -- \\ \hline\hline
\end{tabular}
\caption[]{Predicted fluxes compared to the \eB\ standard solar model for
various combinations of experiments. From~\cite{modind}.}
\label{tab4}
\end{table}

\begin{figure}
\postscript{/home/pgl/fort/nc/graph/solar/ga.ps}{0.8}
\caption[]{Predictions for the gallium rate
of explicit nonstandard models which agree with
Kamiokande, compared with the experimental
observations.  From \protect\cite{seattle,thesis}.}
\label{fig5b}
\end{figure}

\section{Neutrino Oscillations}

Now let's turn to neutrino oscillations.  Suppose that the weak eigenstate
neutrinos, {\it i.e.}, the ones that are produced along with a definite
lepton in weak transitions, are mixtures of the neutrinos of definite mass,
\beqa | \nu_e\rangle &=& | \nu_1 \rangle \cos \theta + |\nu_2 \rangle \sin
\theta \nonumber \\
|\nu_f \rangle &=& - |\nu_1 \rangle \sin \theta + |\nu_2 \rangle \cos
\theta, \label{eq10b} \eeqa
where $\nu_f = \nu_\mu$, $\nu_\tau$, or a sterile neutrino\footnote{ A
sterile neutrino is an $SU_2$ singlet, which has no weak interactions
except by mixing.} $\nu_s$. $\nu_1$ and $\nu_2$ are mass eigenstates and
$\nu_e$ and $\nu_f$ are weak eigenstates.  In a weak decay one produces a
definite weak eigenstate, {\it e.g.},
\beq | \nu (0) \rangle = | \nu_e \rangle .\label{eq11} \eeq
There will then be quantum
mechanical oscillations, just as for any two state system.  At a later
time there will be a probability that the final state
\beqa | \nu (t) \rangle &=& | \nu_1 \rangle e^{-iE_lt} \cos \theta + |
\nu_2 \rangle e^{-iE_2t} \sin \theta \nonumber \\
&=& c_e (t) | \nu_e \rangle + c_f (t) |\nu_f \rangle \neq | \nu_e \rangle
\label{eq12} \eeqa
is different from the initial one.  In particular, there is a
survival probability
\beq P (\nu_e \ra \nu_e; L) = 1 - \sin^2 2\theta \sin^2 \frac{\Delta
m^2L}{4E} \label{eq13} \eeq
of measuring a $\nu_e$ ({\it i.e.}, a probability $1 - P$ of $\nu_e$
disappearance), and a probability
\beq P(\nu_c \ra \nu_f; L) = \sin^2 2 \theta \sin^2 \frac{\Delta m^2L}{4E}
\label{eq14} \eeq
of the appearance of the other neutrino flavor.  Here $\Delta m^2 = m_2^2 -
m_1^2$ is the difference between squared masses, and $m_{1,2} \ll E$ has
been assumed.  The last argument can be written as $1.27 \Delta m^2 L/E$,
where $\Delta m^2$ is in $\rm eV^2$, $L$ is in $m$, and $E$ is in MeV.

\subsection{Mikheyev-Smirnov-Wolfenstein (MSW)}

The above formalism applies to neutrinos passing through vacuum.  In the
presence of matter the neutrinos acquire effective masses from coherent
scattering processes.  In particular, coherent $\nu_ee^- \ra \nu_ee^-$
scattering via the charged current amplitude differentiates the $\nu_e$
from the other neutrinos.  The MSW propagation equation \cite{msw} is
\beqa i \frac{d}{dt}  \left( \begin{array}{c} \nu_e \\ \nu_{\mu,\tau,s}
\end{array} \right) =& \\         \label{eq15}
&\frac{1}{2} \left( \begin{array}{cc}
- \frac{\Delta m^2}{2E} \cos 2 \theta + \sqrt{2} G_Fn &  \frac{\Delta
m^2}{2E} \sin^2\theta \\ \frac{\Delta m^2}{2E} \sin^2 \theta & \frac{\Delta
m^2}{2E} \cos 2 \theta - \sqrt{2} G_Fn \end{array} \right) \left(
\begin{array}{c} \nu_e \\ \nu_{\mu, \tau,s} \end{array} \right),\nonumber
\eeqa
where $G_F \sim 1.2 \x 10^{-5}$~GeV$^{-2}$ is the Fermi constant and
\beqa n = \left\{ \begin{array}{ll} n_e, & {\rm for} \; \nu_e \ra \nu_\mu \;
{\rm or} \; \nu_\tau \\
n_e - \frac{1}{2} n_n, & {\rm for} \; \nu_e \ra \nu_s. \end{array} \right.
\label{eq165}\eeqa
$n_e(n_n)$ is the density of electrons (neutrons). The extra term appears
for sterile neutrinos because of the difference in neutral current
amplitudes.  In the absence of matter, (\ref{eq15}) reproduces
the vacuum oscillation equation.  However, the matter term can be extremely
important.  The MSW resonance occurs for
\beq \frac{\Delta m^2}{2E} \cos 2 \theta = \sqrt{2} G_Fn. \label{eq17} \eeq
Then the diagonal components become equal and
even a  small mixing term  can be amplified to maximal mixing.

If one had a definite density of matter then the resonance would require a
fine-tuning.  However, the neutrinos are born in the center of the sun
where the density is high and as they move outward the density decreases.
For a range of energies (for a given $\Delta m^2$ and $\theta$) a neutrino
will encounter a layer of just the right density for the MSW resonance.  As
it passes through the two energy eigenvalues will ``cross''.  If the density
varies slowly enough (adiabatically) there will be an almost certainty of
conversion.  If the crossing is non-adiabatic, the conversion probability
is smaller.

The conversion probability as the neutrino emerges from the sun depends on
the energy as well as the parameters $\Delta m^2$ and $\sin^2 2 \theta$.
The survival probability is constant (for a fixed energy) within a triangular
region of the $\Delta m^2 - \sin^2 2 \theta$ plane, as shown for the
Kamiokande energies in Figure~\ref{fig6}.  The upper branch of the triangle
is the adiabatic solution \cite{adiabatic}.  On this branch the
density varies adiabatically, and all neutrinos are converted if they
encounter a resonance density.  This occurs for the high energy but not the
low energy neutrinos, so one expects suppression of the high energy part of
the spectrum.  The diagonal, or non-adiabatic, branch (NA) is the one in
which the adiabatic approximation is breaking down \cite{nonadiabatic}.  In
this case the dominant suppression will be in the middle of the spectrum.
Finally, the vertical or large angle branch (LA) is an extension of vacuum
oscillations.  In the regime $\Delta m^2 L/4E \gg 1$ it yields roughly
equal suppression for all energies.

\begin{figure}
\postscript{/home/pgl/fort/nc/graph/solar/Fig5a.ps}{0.6}
\caption[]{Contours of constant survival probability for the Kamiokande
experiment.  The Earth effect is not included.  The contours average over
the relevant neutrino energies.}
\label{fig6}
\end{figure}

Typical probabilities of survival as a function of the neutrino energy for
realistic MSW parameters on the NA and LA branch are indicated in
Figure~\ref{fig7}.
\begin{figure}
\postscript{/home/pgl/fort/nc/graph/solar/Fig8.ps}{0.6}
\caption[]{Survival probabilities for the non-adiabatic and large angle
solutions.}
\label{fig7}
\end{figure}

For several years my collaborators and I have been carrying out the best
MSW analysis of the data that we could \cite{hl2,bhkl}.  We generally use the
Bahcall-Pinsonneault \cite{BPSSM} standard solar model predictions for the
initial fluxes and also for the distributions $\varphi_i(r)$, $n_e(r)$, and
$n_n(r)$, which are respectively the radial distributions of the production
locations of each neutrino flux component and of the electron and neutron
number densities.  We also use other ansatzes for the initial fluxes.  In
analyzing the data one must take into account the energy resolution and
threshold effects for the Kamiokande experiment.  In addition, if the
converted neutrino is a $\nu_\mu$ or a $\nu_\tau$ the neutral current cross
section $\nu_\mu e \ra \nu_\mu e$, which is about $1/6^{\rm th}$ as strong
as the $\nu_e$ charged current cross section, must be included for
Kamiokande.  This effectively lowers the number of surviving $\nu_e$
observed by Kamiokande.

It is important to properly incorporate the theoretical uncertainties
in the initial neutrino fluxes.  These can be due to the core temperature
$T_C$, as well as the production and detector cross sections.  One must
also include the correlations\footnote{Some authors have reached
erroneous conclusions due to the neglect of correlations or the
naive overlapping of contours.}
of those uncertainties between different flux
components and between different experiments \cite{hl2}.  For example, if
the core temperature is higher than in the SSM, it is higher for all of the
flux components and all of the experiments.  To allow for comparison with
updated SSMs and with alternate SSMs we have generally worked with
error matrices parameterized by the temperature and cross
section uncertainties~\cite{bhkl,hl2}.  These are calibrated from
specific Monte Carlos \cite{bu}, and the agreement between the two methods
is excellent, both for the uncertainties and their correlations.
Altogether, the theoretical errors are important but not
dominant.  In analyzing the data it is important to do a joint $\chi^2$
analysis of the data to find the allowed regions.  Simply overlapping
allowed regions between different experiments necessarily neglects
correlations and tends to overestimate the allowed regions.  There are also
complications in the analysis due to the multiple solutions \cite{hl2}.

The Earth effect \cite{earth}, {\it i.e.}, the
regeneration of $\nu_e$ in the Earth at night,  is significant for a
small but important region of the MSW parameters, and not only affects the
time-average rate but can lead to day/night asymmetries.  The Kamiokande
group has looked for such asymmetries and has not observed them
\cite{daynight}, therefore excluding a particular region of the MSW
parameters in a way independent of astrophysical uncertainties.  We fold
both the time-averaged and the day/night data into the overall fits
\cite{hl1,hl2}.

\begin{figure}
\postscript{/home/pgl/fort/nc/graph/solar/p_comb_bpssm_0694.ps}{0.8}
\caption[]{Allowed regions at 95\% CL from individual experiments and from
the global fit.  The Earth effect is included for both time-averaged and
day/night asymmetry data, full astrophysical and nuclear physics
uncertainties and their correlations are accounted for, and a joint
statistical analysis is carried out.  The region excluded by the Kamiokande
absence of the day/night effect is also indicated. From
\protect\cite{hl2,thesis}.}
\label{fig8}
\end{figure}

The allowed regions from the overall fit for normal oscillations
$\nu_e \ra \nu_\mu$ or $\nu_\tau$ are shown in Figure~\ref{fig8}.  There
are two solutions at 95\% C.L., one in the NA branch for the Homestake and
Kamiokande experiments (and the adiabatic branch for the gallium
experiments), and one on the LA branch.  The former gives a much better
fit, as can be seen in Table~\ref{tab5}.  There is a second large angle
solution with smaller $\Delta m^2$, which only occurs at 99\% C.L.

\begin{table} \centering
\begin{tabular}{|lccc|} \hline \hline
                & Non-adiabatic &     Large Angle I    &  Large Angle II \\
\hline
$\sin^2 2 \theta$& $6.5 \x 10^{-3}$  & $0.62$ & $0.76$ \\
$\Delta m^2 ({\rm eV}^2)$ & $ 6.1 \x 10^{-6}$ & $9.4 \x 10^{-6}$ & $1.2 \x
10^{-7}$ \\ \hline
$\chi^2 (7df)$ & $3.1$ & $8.1$ & $13.1$ \\
$P  (\%)$   & 88 & 32 & 7 \\
$P_{\rm relative}$(\%) & 94.9 & 4.6 & 0.5 \\ \hline \hline
\end{tabular}
\caption[]{MSW parameters for the non-adiabatic and large angle solutions as
well as the overall $\chi^2$ (7df).  There is also a second large angle
solution, which is allowed at 99\% C.L. but
gives a much poorer fit.  The last
two rows are the probability in each case of obtaining  a larger $\chi^2$,
and the relative probabilities of the various solutions. From
\protect\cite{hl2,thesis}.}
\label{tab5}
\end{table}

MSW fits can also be performed using other solar models as inputs, as a way
of getting a feeling for the uncertainties.  Figure~\ref{fig9} shows the
MSW fit assuming the  TCL SSM \cite{TCSSM}.  One sees that the allowed
regions are qualitatively similar, but differ in detail.

\begin{figure}
\postscript{/home/pgl/fort/nc/graph/solar/p_comb_tcssm.ps}{0.8}
\caption[]{Allowed regions assuming the TCL SSM. From
\protect\cite{hl2,thesis}.}
\label{fig9}
\end{figure}

One can also consider transitions $\nu_e \ra \nu_s$ into
sterile neutrinos.  These are different in part because the MSW formulas
contain a small contribution from the neutral current scattering from
neutrons.  Much more important is the lack of the neutral current
scattering of the $\nu_s$ in the Kamiokande experiment.  There is a
non-adiabatic solution similar to the one for active neutrinos, though the
fit is poorer.  However, there is no acceptable large angle solution
because of the lack of a neutral current, which makes that case similar to
astrophysical solutions.  Oscillations into a sterile neutrino in that
region are also disfavored by Big Bang nucleosynthesis \cite{sterile}.

It is interesting to go a step further and consider nonstandard solar
models and MSW simultaneously \cite{bhkl,hl2,modind}.  There is now
sufficient data to determine both the MSW parameters and the core
temperature in a simultaneous fit.  One obtains \cite{modind,thesis}, $T_C =
1.00 \pm 0.03$, in remarkable agreement with the standard solar model
prediction $1 \pm 0.006$.  The allowed MSW parameters are shown in
Figure~\ref{fig10}.  The regions are larger than when one accepts the
SSM, but still constrained.

\begin{figure}
\postscript{/home/pgl/fort/nc/graph/solar/p_comb_tcfree.ps}{0.8}
\caption[]{Allowed regions of the MSW parameters when $T_C$ is allowed to be
free. From \protect\cite{modind,thesis}.}
\label{fig10}
\end{figure}

\begin{figure}
\postscript{/home/pgl/fort/nc/graph/solar/p_comb_Bfree.ps}{0.8}
\caption[]{Allowed MSW parameters when the \eB\ flux is  free. From
\protect\cite{modind,thesis}.}
\label{fig11}
\end{figure}

Alternatively, one can allow the \eB\ flux $\phi (\,^8B)$ to be free, as
would be expected in models with lower $S_{17}$, for example.  The data are
consistent with the SSM value with large errors, but favor a slightly
higher value $\phi(\,^8B)/\phi_{SSM} = 1.15 \pm 0.53$.  The allowed regions
of the MSW parameters are shown in Figure~\ref{fig11}.

Although the MSW  mechanism gives a perfect description of
existing data, there is one
alternative, vacuum oscillations \cite{vac1}--\cite{h1}.  There are fine-tuned
solutions with the earth-sun distance being at a node of the oscillations,
corresponding to parameter ranges $\Delta m^2 \sim 10^{-10} \ev^2$ and
$\sin^2 2\theta > 0.7$.

\section{The Future}

In the future one will want to verify or falsify that MSW is occurring, or
choose some other possibility.  Assuming that MSW is correct, one will want
to find the unique solution for the parameters and tell whether the
oscillations are into active or sterile neutrinos.  Not only will the
gallium experiments have better statistics and calibrations but, in
addition, there will be a new generation of experiments.  For example, the
Sudbury Neutrino Observatory (SNO), which is a heavy water experiment being
built by a Canadian-American collaboration in Canada, will have many new
capabilities.  It will be able to measure charged current reactions from
deuterium,
\beq \nu_e + d \ra p + p + e^- . \label{eq18} \eeq
It will also be able to measure the neutral current reaction,
\beq \nu_{e,\mu,\tau} + d \ra \nu_{e,\mu,\tau} + p + n, \label{eq19} \eeq
and will observe       electron scattering
\beq \nu_{e,\mu,\tau} + e \ra \nu_{e,\mu, \tau} + e   \label{eq20} \eeq
with both charged and neutral current amplitudes.  The ratio of neutral
current to charged current rates can be measured directly or by comparing
the charged current rate with the electron scattering results from SNO or
from the Super-Kamiokande experiment (a large water Cerenkov experiment
being built in Japan).  If different from the weak interaction
expectations, the ratio will clearly indicate neutrino oscillations, either
vacuum or MSW. Therefore, one should be able to distinguish new neutrino
physics\footnote{Oscillations into sterile neutrinos would not affect the
neutral current to charged current ratio, but could still yield spectral
distortions or day/night asymmetries.} from astrophysical solutions
in a way independent of astrophysical uncertainties.  Assuming this is
observed, one will be able to use the neutral current events to calibrate the
initial \eB\ flux and determine the core temperature.  (Or, more
accurately, a combination of the core temperature and $S_{17}$.)  In
addition, the SNO experiment will be sensitive to spectral distortions.
These are expected to be significant for the non-adiabatic MSW solution,
but not for the large angle, and are essentially free of astrophysical
uncertainties.  Vacuum oscillation solutions would also give
large spectral distortions.  This is illustrated in Figure~\ref{fig12}.
The Super-Kamiokande experiment will have  high
statistics for  electron scattering and should also be able to observe
spectrum distortions.

\begin{figure}
\postscript{/home/pgl/fort/nc/graph/solar/fig5.ps}{0.6}  
\caption[]{Expected spectrum distortions for the MSW non-adiabatic and
various vacuum oscillation experiments, with expected uncertainties.  The
large angle solution exhibits no significant distortions and is similar to
an astrophysical solution. From \protect\cite{h1,seattle,thesis}.}
\label{fig12}
\end{figure}

\begin{figure}
\postscript{/home/pgl/fort/nc/graph/solar/fig8a.ps}{0.65}
\postscript{/home/pgl/fort/nc/graph/solar/fig8b.ps}{0.65}
\postscript{/home/pgl/fort/nc/graph/solar/fig8c.ps}{0.65}
\caption[]{Expected seasonal variations for various vacuum oscillation
solutions   in the SNO, Super-Kamiokande, and BOREXINO experiments. From
\protect\cite{h1,seattle,thesis}.}
\label{fig13}
\end{figure}

Yet another probe that is free of astrophysical uncertainties is that the
large angle solution is expected to yield a significant day/night asymmetry
in the SNO and Super-Kamiokande experiments.  This may even be observable
in Kamiokande~III.  Not only are there the day/night effects expected
from MSW, but one expects large seasonal variations for vacuum
oscillations, as indicated in Figure~\ref{fig13}.

Another experiment, BOREXINO in the Gran Sasso Laboratory in Italy, will be
sensitive to the \sBe\ line.  This is especially interesting because there
is every indication that this is where the dominant suppression occurs.
Furthermore, other explanations of the solar neutrino problem, such as
large magnetic moments \cite{mag}, would imply interesting consequences for
the \sBe\ line.

Altogether, these experiments should be able to establish or refute the MSW
solution, determine the parameters, and probe alternatives precisely.
Whatever happens, one is still interested in the neutrinos not only for
particle physics but also for astrophysics.  Fortunately, even if MSW is
going on it should be possible to establish it and constrain the parameters
from the methods described above, with little uncertainty from astrophysics
or nuclear physics.  In fact, there should be enough data not only to
determine the MSW parameters but also to determine the initial $pp$, $\!^8B$,
and \sBe\ flux components in a model independent way, much as was described
in Section~\ref{modindsec} for astrophysical solutions without MSW
\cite{modind}.  For this program one will need, in addition to SNO and
Super-Kamiokande, the BOREXINO measurement of the \sBe\ line.  That is
because MSW can lead to an almost total suppression of the \sBe $\nu_e$
flux.  However, BOREXINO will have a neutral current sensitivity to the
converted neutrinos at about 20\% of the $\nu_e$ efficiency,
yielding a measure of the
initial flux.  In addition to the model independent studies, one can
determine the parameters of the standard solar models, such as $T_C$ and
$S_{17}$, simultaneously with $\Delta m^2$ and $\sin^ 2 \theta$.  The
projected sensitivity is shown in Figure~\ref{fig14}.

\begin{figure}
\postscript{/home/pgl/fort/nc/graph/solar/tb_90cl.ps}{0.6}
\caption[]{Projected uncertainties in the simultaneous determination of
$T_C$, $S_{17}$, and the MSW parameters, assuming future experiments. From
\protect\cite{modind}.}
\label{fig14}
\end{figure}

\section{Implications}

There are literally hundreds of models of neutrino mass \cite{nurev}.
However, many theories with coupling constant unification, such as grand
unified theories, predict a seesaw-type mass \cite{seesaw,BKL}
\beq m_{\nu_i} \sim \frac{C_i m^2_{u_i}}{M_N}, \label{eq22} \eeq
where $M_N$ is the mass of a superheavy neutrino, $u_i = u$, $c$, $t$ are
the up-type quarks, and $C_i$ is a radiative correction.  The general
$\Delta m^2$ range suggested by the solar neutrinos is consistent with the
GUT-seesaw range.  In particular, in the string motivated models one
expects the heavy mass to be a few orders of magnitude below the
unification scale \cite{cl}.  As an example, for $M_N \sim 10^{-4} M_{GUT}
\sim 10^{12}$~GeV one predicts
\beqa m_{\nu_e} &<& 10^{-7} \ev \nonumber \\
m_{\nu_\mu} & \sim & 10^{-3} \ev \label{eq23} \\
m_{\nu_\tau} &\sim& (3 - 21) \ev.\nonumber \eeqa
In this case one would expect oscillation of $\nu_e \ra \nu_\mu$ in the
sun, and perhaps the $\nu_\tau$ is in the range relevant to hot dark
matter.  If this is the case there is a good chance that $\nu_\mu \ra
\nu_\tau$ oscillations will be observed in accelerator appearance
experiments now
underway at CERN.  Alternately, for small modifications in the seesaw one
could have somewhat smaller $\nu_\tau$ masses that could lead to $\nu_\mu
\ra \nu_\tau$ oscillations in the range relevant to the atmospheric
neutrino anomaly \cite{atmun}.

The specific predictions are highly model dependent, and one cannot make
anything more than general statements at this time.  It will be important
to follow up all experimental possibilities.  If oscillations are
responsible for the atmospheric neutrino results it should possible to
prove it with long baseline oscillation experiments proposed at Fermilab
and Brookhaven.

It is difficult to account for solar neutrinos, a component of hot dark
matter, and atmospheric neutrinos simultaneously.  There are just not
enough neutrinos to go around.  Attempts to account for all of these
effects must invoke additional sterile neutrinos and/or nearly
degenerate neutrinos, so that the mass differences can be much smaller than
the average masses \cite{caldmoh}.

\section{Conclusions}

\begin{itemize}

\item
There is no compelling laboratory evidence for neutrino mass.
\item  However, neutrino masses are non-zero in most extensions of the
standard model, typically             $m_\nu \sim v^2/M_{new}$,
where $v = 246$~GeV is the weak scale.  Thus, probing small neutrino masses
indirectly probes the large mass scales of new physics.

\item A deficit of solar neutrinos is observed in four experiments.  The
less serious solar neutrino problem is the deficit of \eB\ neutrinos, which
could well be accounted for by small changes in the astrophysics.  Much
more serious is that any two of the three classes of experiments can be
combined to indicate that the dominant suppression is in the middle of the
spectrum, in particular of the \sBe\ neutrinos and the lower energy part of
the \eB\ neutrinos.  This is incompatible with any known astrophysical or
nuclear physics explanation, suggesting either non-standard neutrino
properties or that some of the experiments are wrong.  On the other hand,
the MSW oscillations give a perfect description of the data.  Detailed MSW
analyses, including the Earth effect, the day/night asymmetry data,
theoretical uncertainties, and their correlations indicate that there are
two parameter solutions, both in the general range expected from grand
unification, although the details are model dependent.  One can also
simultaneously determine the MSW and solar parameters.  One finds a core
temperature $T_C/T_{SSM} = 1.00 \pm 0.03$, in remarkable agreement with the
standard model expectation $1.0 \pm 0.006$.  Alternately, one can constrain
the boron neutrino flux, yielding $\phi (\,^8B)/\phi_{SSM} = 1.15 \pm 0.53$,
consistent with but slightly higher than the SSM expectation.  In the
future it should be possible with new experiments to determine the initial
$pp$, \sBe, and \eB\ fluxes, with or without MSW being present, in a model
independent way.

\item There are also anomalies in the atmospheric $\nu_\mu/\nu_e$
ratio produced by cosmic ray interactions in the atmosphere.  This could
also     be a sign of neutrino oscillations.

\item The combination of COBE data and the large and small scale
distribution of galaxies is hard to account for by pure cold dark matter.
One possibility is a component of hot dark matter, such as a $\tau$
neutrino in the 10~eV range.

\item The possibilities are very exciting.  It is quite possible that
neutrino mass will be the first clear break with the standard electroweak
model.  However, much experimental and theoretical work remains.

\end{itemize}

\section*{Acknowledgement}

It is a pleasure to thank Naoya Hata for collaboration in this work.

\end{document}